# Gender Differences in Undergraduate Physics Courses: A Comparative Study of Persistence

Ashiqul Islam Dip[1,*], Mohammad Tomal Hossain[2], Md Forman Ullah[3], Md Salah Uddin[2]

[1]Division of Experimentation, Community of Physics, Dhaka, Bangladesh
[2]Division of Administration, Community of Physics, Dhaka, Bangladesh
[3]Division of Academics, Community of Physics, Dhaka, Bangladesh

**Email Address:**
dip@communityofphysics.org (Ashiqul Islam Dip), tomal@communityofphysics.org (Mohammad Tomal Hossain), forman@communityofphysics.org (Md Forman Ullah), salahuddin@communityofphysics.org (Md Salah Uddin)
[*]Corresponding Author

**Abstract:** We have investigated the difference in persistence between male and female students while taking undergraduate physics courses. To quantify the persistence of a certain group of students, we have defined 'persistence index' as the inverse of the decrease rate of the number of that group of students while taking a specific course. We have collected the data from three consecutive workshops on various topics of physics. After plotting the number of participations against the number of days attended, we have calculated the decrease rates and persistence indices for both male and female student groups on each workshop and compared the persistence indices on a bar diagram. The comparative statistics show that the persistence indices of female student groups are significantly higher than that of male student groups. This leads us to the conclusion that the female students are more persistent than male students while taking an undergraduate physics course.

**Keywords:** Physics Education, Undergraduate Physics Course, Gender Difference, Persistence Index

## 1. INTRODUCTION

It is a well-documented fact that the disciplines of science, technology, engineering and math (STEM) are predominated by male students. Agreeing to some studies, women in physics only comprise approximately 19% of all undergraduate and graduate students [1-5]. Some other studies indicate that women show lower levels of conceptual knowledge than men in both beginning and ending of introductory physics courses [6,7]. According to the research of Kost *et al.*, women show less involvement in learning and problem solving [7]. These gender differences increase, for both conceptual knowledge and involvement, along with the evolution of the course [6,7]. Kost-Smith *et al.* [8] found that women entered introductory physics courses with lower self-efficacy than men, and this disparity also increased along with the development of the course. In lecture-based physics courses, Sawtelle *et al.* [9] obtained the same result, as did Cavallo *et al.* [10], and Lindstrom and Sharma [11]. Another investigation by Kost-Smith *et al.* [12] says, women exhibit less expert attitude than men.

As a part of a non-profit educational institution, Community of Physics[†], we have organized and conducted several workshops focusing on diverse topics in physics and mathematics. In the beginning of every workshop, we have seen that the male participants exceptionally outnumbered the female participants. After the first day, the numbers of the participants in both groups start to decline, and on the last day, the number of female students and the number of male students become nearly equal. This consistent behavior of the students piqued our curiosity and lead us to hypothesize that the female students show a higher persistence than the male students. Thus, we were inclined to conduct a formal research to check the validity of our hypothesis.

## 2. METHODS

### 2.1. COLLECTION OF DATA

We administered our study in three workshops. Each of the workshops explored the physical and mathematical aspects of a distinct field of interest. The first one was on vector calculus, the second one was on Newtonian mechanics, and the third one was on classical electromagnetism. All the participants were undergraduate students of various disciplines of physical sciences and engineering from several Bangladeshi universities. Participation data were collected on a daily basis.

---

[†] http://www.communityofphysics.org

The inaugural workshop was labeled as *1st Workshop on Vector Calculus (WVC1)*. The workshop covered vector algebra, single-variable differential and integral calculus, multi-variable and parametric functions, partial derivatives, multi-variable integral calculus, fundamental theorems of vector calculus, vector analysis on curved manifolds, Cartesian tensors and Maxwell's equations as an application of vector calculus. We prepared course materials following *Calculus* by Anton *et al.* [13], *Calculus* by Strauss *et al.* [14], Banchoff, and Lovett's *Differential Geometry of Curves and Surfaces* [15], and *Vector Analysis* by Spiegel and Lipschutz [16]. *WVC1* was a six-day workshop. It ran for six days starting from 8:30 am to 5:00 pm with a one-hour break.

On the starting day, there were 89 participants, of whom 66 were male and 17 were female. Of the total 89 participants, 21 (25.3%) were from physics, 13 (15.7%) were from mathematics, 20 (24.1%) were from electrical engineering, 9 (10.8%) were from computer science, 6 (7.2%) were from communication engineering, 4 (4.8%) were from civil engineering, and 11 (13.3%) were from mechanical and other engineering disciplines.

The epithet of the second workshop was *1st Workshop on Classical Mechanics: From Newton to Lagrange (WCM1)*. This workshop covered preliminary mathematical tools, Newton's laws, projectile motion, drag force, conservation of momentum, conservation of energy, oscillation (simple, damped & damped-driven), Newtonian gravity, Kepler's laws, mechanics in non-inertial frames, D'Alembert's principle and Lagrange's equation. The course materials were prepared using Jefferson and Beadsworth's *Further Mechanics* [17], *Introduction to Classical Mechanics: With Problems and Solutions* by Morin [18], *Classical Mechanics* by Goldstein *et al.* [19], and Taylor's *Classical Mechanics* [20]. It was a five-day workshop. Starting from 9:00 am, the workshop ran up to 5:30 pm with a one-hour break in between.

On the starting day of the workshop, there were 57 participants, of whom 47 were male and 10 were female. Of the total 57 participants, 15 (26.3%) were from physics, 8 (14.0%) were from mathematics, 5 (8.8%) were from chemistry, 11 (19.3%) were from electrical engineering, 6 (10.5%) were from computer science, 5 (8.8%) were from mechanical engineering, and 4 (7.0%) were other engineering students.

The third workshop was termed as *1st Workshop on Classical Electromagnetism (WEM1)*. The workshop explored vector analysis, Helmholtz theorem, electrostatic field equations, electrostatic force and energy, Poisson's and Laplace's equation, Green's function, polarization, dielectric medium, magnetostatic field equations, magnetostatic force and energy, magnetization, magnetic materials, Maxwell's equations, conservation laws in electromagnetism, potential formulation, electromagnetic waves and special relativity. We used Griffiths' *Introduction to Electrodynamics* [21], Zangwill's *Modern Electrodynamics* [22], and Franklin's *Classical Electromagnetism* [23] in preparing course materials. With a one-hour break, the workshop ran from 8:00 am to 6:00 pm for five days.

On the starting day, there were 38 participants, of whom 28 were male and 10 were female. Of the total 38 participants, 12 (31.6%) were from physics, 3 (7.9%) were from chemistry, 2 (5.3%) were from mathematics, 9 (23.7%) were from electrical engineering, 4 (10.5%) were from mechanical engineering, 3 (7.9%) were from computer science, 2 (5.3%) were from chemical engineering, and 3 (5.3%) were from other engineering disciplines.

As our data represent a diverse body of students of a multitude of backgrounds, it is highly likely, that our data is little touched by random fluctuation. Participants accepted to fill a form by themselves to confirm their presence on the first day of the workshop. On the rest of days of the workshops, they had to sign in the attendance books.

**2.2. AN EMPIRICAL RELATION**

Starting from the first day, the number of the participants decreased in all workshops for both male and female groups. Using the analogy with many decay processes in nature, we have assumed that the number of the participants decreases exponentially with time. If there are $N_0$ number of participants in a specific group at the beginning of a workshop, after time, $t$ (in days), the number of the participants, $N$ is given by the relation

$$N = N_0 e^{-\lambda t}$$

where, $\lambda$ is the decrease rate of the number of the participants of that group. The higher value of $\lambda$ indicates the lower value of persistence.

To quantify persistence, we define the 'persistence index' $\wp$, by the relation

$$\wp = 1/\lambda$$

The higher value of persistence index, $\wp$ of a group is the signature of having higher persistence of that group. However, the persistence index is analogous to the *time constant* of a decay process, which measures how much time it will take to decrease the number of the participants of a group to 36.8% of its initial value.

**2.3. ANALYSIS OF DATA**

The collected data were first tabulated in a spreadsheet using Microsoft Excel 2016. Then the data were sorted to identify the number of the participants of different backgrounds, as presented in the previous section. Further sorting was carried out to separate male and female participants. For each workshop, a set of male and female participation data were obtained.

Then we have plot the number of the participants attending against the number of days for each set of data, and fit each with exponential curves. All the curves had a negative valued exponent showing a decay-like behavior. Thus, from each workshop data, we have obtained one decrease rate for male participants and another decrease rate for female participants. From the obtained decrease rates, we have calculated the persistence indices for each group. Then we have compared the persistence indices of male and female participants in a single bar diagram for each workshop. We have also compared the mean persistence indices of male and female groups. The threshold of significance was set at 0.03.

## 3. DATA AND RESULTS

The result of our investigation is presented for each workshop individually in *FIG. 1-FIG. 6,* and then the combined comparison of persistence indices is presented in *FIG. 7*. In all the plots (*FIG. 1-FIG. 6)*, data points were fitted with exponential curves and the decrease rates were obtained. The mean persistence index for both male and female group, and the level of significance (*p*-value) were calculated.

### 3.1. 1ST WORKSHOP ON VECTOR CALCULUS (WVC1)

In *WVC1*, there were 66 male participants who were present for at least 1 day and there were only 2 male participants with sustained participation for all the six days of the workshop. *FIG. 1* shows the number of the male participants against the number of days attended. The obtained decrease rate is $\lambda = 0.642$. And the obtained persistence index is $\wp = 1.558$[‡].

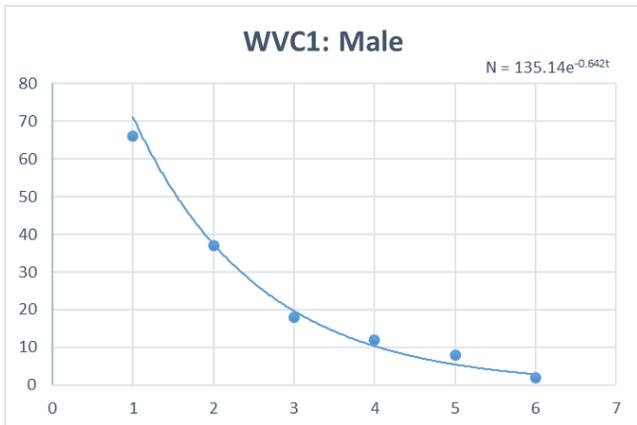

FIG. 1. Number of participants vs. number of days attended for male participants in 1st Workshop on Vector Calculus (WVC1).

On the other hand, there were 17 females who participated at least 1 day and there were only 3 female participants who persisted through all six days of the workshop. *FIG. 2* shows the number of the female participants against the number of days attended. The obtained decrease rate is $\lambda = 0.320$, and the persistence index $\wp = 3.125$[‡]. This persistence index of the female participants is 100.6% higher than the persistence index of the male participants of the same workshop.

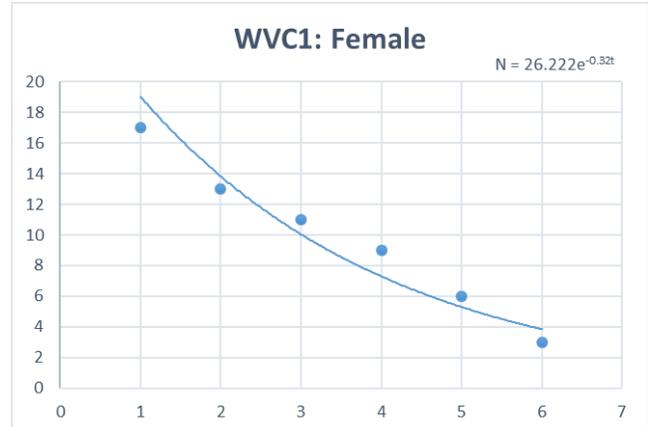

FIG. 2. Number of participants vs. number of days attended for female participants in 1st Workshop on Vector Calculus (WVC1).

### 3.2. 1ST WORKSHOP ON CLASSICAL MECHANICS: FROM NEWTON TO LAGRANGE (WCM1)

In *WCM1*, there were 47 male participants who stayed for at least 1 day and there were 8 male participants who continued through all five days of the workshop. *FIG. 3* shows the number of the male participants against the number of days attended. The obtained decrease rate is $\lambda = 0.442$, and the persistence index is $\wp = 2.262$[‡].

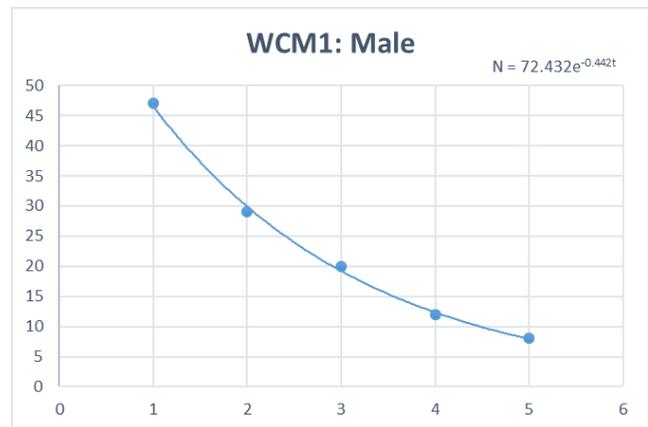

FIG. 3. Number of participants vs. number of days attended for male participants in 1st Workshop on Classical Mechanics: From Newton to Lagrange (WCM1).

In contrast, there were 10 female participants who participated for at least 1 day and there were only 3 female participants who joined us on all the five days of the workshop. *FIG. 4* shows the number of the female participants against the number of days attended. The obtained decrease rate is $\lambda = 0.310$. And the obtained persistence index is $\wp = 3.226^{\ddagger}$. This persistence index of the female participants is 42.6% higher than the persistence index of the male participants of the same workshop.

---

[‡] Correct to three decimal places.

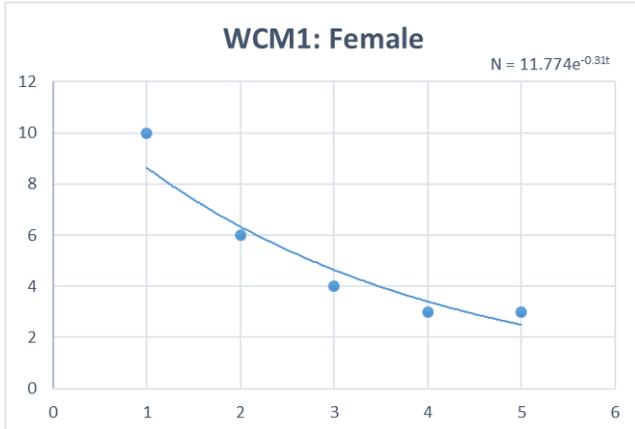

FIG. 4. Number of participants vs. number of days attended for female participants in 1st Workshop on Classical Mechanics: From Newton to Lagrange (WCM1).

### 3.3. 1ST WORKSHOP ON CLASSICAL ELECTROMAGNETISM (WEM1)

In *WEM1*, 28 male participants were present for at least a day, while only 6 males could sustain their interest throughout the five days of the workshop. *FIG. 5* shows the number of the male participants against the number of days attended. The obtained decrease rate is $\lambda = 0.377$, and the persistence index is $\wp = 2.653^{\ddagger}$.

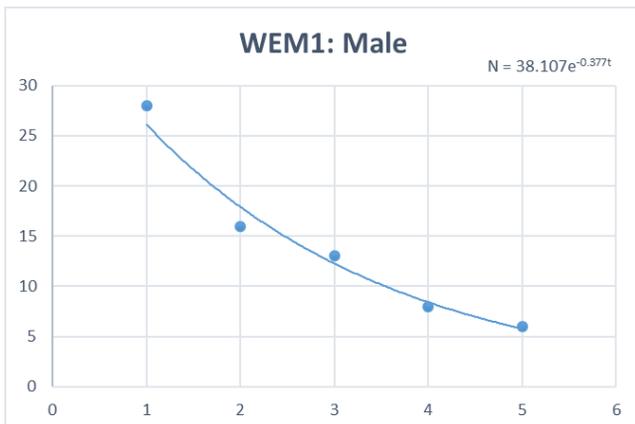

FIG. 5. Number of participants vs. number of days attended for male participants in 1st Workshop on Classical Electromagnetism (WEM1).

Then again, we had a population of 10 females on the first day, which dwindled to 3 over the course of the workshop. *FIG. 6* shows the number of the female participants against the number of days attended. The obtained decrease rate is $\lambda = 0.297$, and the persistence index is $\wp = 3.367^{\ddagger}$. This persistence index of the female participants is 26.9% higher than the persistence index of the male participants in the same workshop.

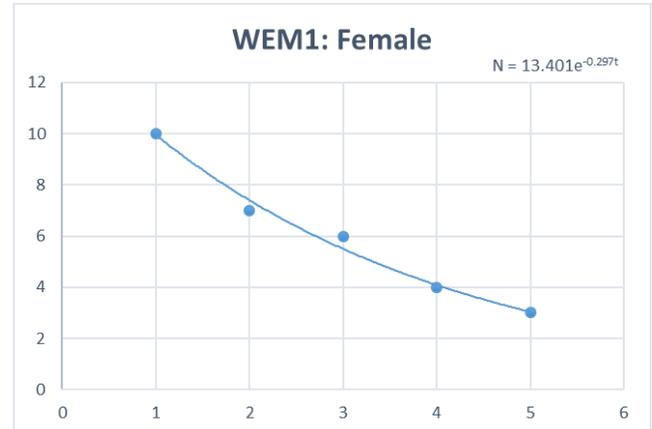

FIG. 6. Number of participants vs. number of days attended for female participants in 1st Workshop on Classical Electromagnetism (WEM1).

### 3.4. COMPARISON BETWEEN THE PERSISTENCE INDICES OF MALE AND FEMALE PARTICIPANT GROUPS

In *FIG. 7*, we show the comparison between the persistence indices of male and female participants graphically. It shows that in each workshop, the persistence index of the female participants is significantly higher than that of the male participants at a tolerance of 25%.

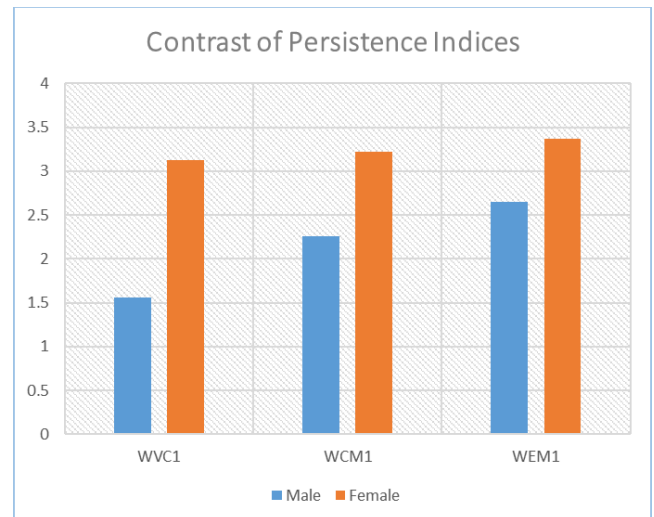

FIG. 7. Contrast of persistence indices for male and female participant groups for each workshop.

### 3.5. FURTHER STATISTICAL ANALYSIS

The mean (average) persistence index, for the male participant groups, is $\bar{\wp} = 2.158$, with a standard deviation $S = 0.555$. On the other hand, the mean persistence index for the female participant groups is $\bar{\wp} = 3.239$, with a standard deviation $S = 0.122$. This statistical comparison tells that the female participants are more persistent than the male participants.

To calculate the confidence interval and $p$-value, we have used $t$-distribution, as our sample size is small. The degrees of freedom of our data is 2. The mean difference of our paired sample is $\bar{d} = 1.082$, with a standard deviation $S_d = 0.439$. The $t$-score of our paired sample is

$$t = \frac{\bar{d}}{\sqrt{\frac{S_d^2}{n}}} = 4.269$$

For this obtained value of $t$-score, the $p$-value is $p = 0.0254$. Thus, our result, that the female participants are more persistent than males, is significant at $p < 0.03$ and has 97% confidence interval.

### 4. DISCUSSION AND CONCLUSIONS

In our study, the three workshops exhibited different values of the persistence indices for both male and female participants. While the persistence indices of the male participants varied wildly ($S = 0.555$) in three workshops, the persistence indices of the female participants remained almost stable ($S = 0.122$). The persistence indices could vary due to other external factors [24], like transportation facility to the location or the workload in the workshop. But the key point is to notice that the persistence indices of the female participants always exceeded the persistence indices of the male participants in a single workshop. The mean persistence index of the female participants is greater than the mean persistence index of the male participants, which is statistically significant ($p < 0.03$). Therefore, we conclude that female students are more persistent than male students while taking a physics course. Previously, McCormick *et al.* had produced similar results in their study [25].

Another important point is that as the workload in the three consecutive workshops gradually increased, persistence indices for both male and female participant groups increased. It could be possible that the increased amount of workload increases the persistence and reduces gender gap, but it requires further study to be proven.

### ACKNOWLEDGEMENTS

We gratefully acknowledge the support of the Community of Physics that has supported the research financially, and provided the necessary data from the conducted workshops. We sincerely thank Dr. Sabina Hussain, Dr. Khandker S. Hossain, Liana Islam, and Md. Arafat Hossen, who spent their valuable time in reviewing this article.